\documentclass[runningheads]{llncs}
\usepackage[T1]{fontenc}
\usepackage{graphicx}
\usepackage{booktabs}
\usepackage[misc]{ifsym}
\usepackage{multirow}
\usepackage{amssymb}
\usepackage{amsmath}
\usepackage{bbm}
\usepackage{slashbox}
\usepackage{enumitem}
\usepackage{xcolor}
\usepackage[hyphens]{url}
\usepackage{tikz}
\usepackage{algorithm}
\usepackage{algpseudocode}
\usepackage{booktabs}
\usepackage{pifont}
\usepackage{subfigure}
\usepackage{adjustbox}
\newcommand{\cmark}{\ding{51}}
\newcommand{\xmark}{\ding{55}}
\graphicspath{{fig/}}
\begin{document}

\title{Kolmogorov-Arnold Network for Gene Regulatory Network Inference}

\titlerunning{scKAN: Kolmogorov-Arnold Network for GRN Inference}
% If the full title of your paper is short enough to also fit in the running head, you can omit the abbreviated paper title here. You can check as follows: if you comment out the \titlerunning line, something will appear in the header of all odd-numbered pages of your PDF from page 3 onward. This something is either the full title (in which case all is well), or the error message "Title Suppressed Due to Excessive Length". If this error message appears, you're going to want to provide an abbreviated title within the \titlerunning command, because if you won't do it, Springer will do it for you.

%N.B.: Author information (both in the \author{} and \authorrunning{} command) should only be present in the Camera-Ready Version of your paper. The version that you initially submit for review, ought to be double-blind. So, when initially submitting your paper, use:
%\author{Author information scrubbed for double-blind reviewing}
\author{
Tsz Pan Tong\inst{1,2}\orcidID{0000-0001-8111-5886} \and
Aoran Wang\inst{1}\orcidID{0000-0001-7809-0622} \and
George Panagopoulos\inst{1}\orcidID{0000-0001-7731-9448} \and
Jun Pang\inst{1,2}\orcidID{0000-0002-4521-4112}
}
% You may leave out the orcidID information, if you want to.
% Use \corr to indicate the corresponding author. Note the spacing around the \corr command. Only one author can be the corresponding author.

%N.B.: comment out the \authorrunning{} command for the double-blind version of your paper submitted for review. Later, if your paper is accepted, use the command for the Camera-Ready Version.
\authorrunning{T.~P. Tong et al.}
% If there are two authors, write 'A.L. Benjamin and C.C. Broadus Jr.'
% If there are more than two authors, '[...] et al.' is used.

\institute{Department of Computer Science, University of Luxembourg, Luxembourg \and Institute for Advanced Studies, University of Luxembourg, Luxembourg \\
\email{\{tszpan.tong,aoran.wang,georgios.panagopoulos,jun.pang\}@uni.lu}}

\maketitle              % typeset the header of the contribution

\begin{abstract}
    Gene regulation is central to understanding cellular processes and development, potentially leading to the discovery of new treatments for diseases and personalized medicine.
    Inferring gene regulatory networks (GRNs) from single-cell RNA sequencing (scRNA-seq) data presents significant challenges due to its high dimensionality and complexity.
    Existing tree-based models, such as GENIE3 and GRNBOOST2, demonstrated scalability and explainability in GRN inference, but they cannot distinguish regulation types nor effectively capture continuous cellular dynamics.
    In this paper, we introduce scKAN, a novel model that employs a Kolmogorov-Arnold network (KAN) with explainable AI to infer GRNs from scRNA-seq data.
    By modeling gene expression as differentiable functions matching the smooth nature of cellular dynamics, scKAN can accurately and precisely detect activation and inhibition regulations through explainable AI and geometric tools.
    We conducted extensive experiments on the BEELINE benchmark, and scKAN surpasses and improves the leading signed GRN inference models ranging from 5.40\% to 28.37\% in AUROC and from 1.97\% to 40.45\% in AUPRC.
    These results highlight the potential of scKAN in capturing the underlying biological processes in gene regulation without prior knowledge of the graph structure.

\end{abstract}

%------------------------------------------
\section{Introduction}
\label{sec:intro}
%------------------------------------------
Genes hold the key information needed for the growth and functioning of living organisms.
Gene regulation involves controlling the expression of genes that are vital for cell growth, development, and maintenance.
Thus, reconstructing gene regulatory networks (GRNs) is essential to uncover the mechanisms behind biological processes and diseases and to promote the development of precision medicine~\cite{van2018integrative} and personalized treatments~\cite{burska2014gene,cooper2016hypertension}.
With the advance of sequencing technology, single-cell RNA sequencing (scRNA-seq) has made it possible to capture the gene expression distribution across individual cells at discrete sampling points.
The high data volume of the scRNA-seq assay provides fertile soil for the development of sophisticated and precise GRN models.

GENIE3~\cite{huynh2010genie3} and GRNBOOST2~\cite{moerman2019grnboost2} are two remarkable state-of-the-art tree-based GRN inference models in the era of scRNA sequencing.
They are recognized as the most robust and recommended GRN inference models in the BEELINE benchmark~\cite{pratapa2020benchmarking}, and were integrated into the popular SCENIC+ pipeline~\cite{aibar2017scenic,bravo2023scenic+}.
Their success can be attributed to the One-vs-Rest (OvR) formulation, where each gene is modeled and reconstructed as a function of other genes.
This formulation enables model scalability to thousands of genes and explainability through the learned importance scores in the tree models.

Despite their effectiveness, GENIE3 and GRNBOOST2 are tree models, which inherently introduce discontinuities in reconstructed gene expressions due to stacked decision boundaries.
At the molecular level, the ligand-receptor coupling and transcription factor binding are discrete events that occur in seconds, which can be modeled by chemical master equations (CMEs)~\cite{liang2010computational,qian2010chemical,gorin2022modeling}.
However, the majority of the measurement resolution is limited to hours or days at the cellular level, and all stochastic events are averaged, exhibiting a continuous process.
Thus, using ODEs that match continuous dynamics is a more common choice~\cite {matsumoto2017scode,tong2020trajectorynet,sha2024tigon}.
In addition, neither model can distinguish between the activation and inhibition of the regulations, limiting their application in the biological context.
Moreover, both models give an averaged regulatory strength, which ignores differences in cell lineages and buries signals from rare cell types~\cite{wu2021bayesian}.

Motivated by this, we propose a novel model, scKAN, to infer GRNs from scRNA-seq data with Kolmogorov-Arnold networks (KAN)~\cite{liu2024kan}.
KAN can guarantee a continuous model and has shown promising results in regression tasks with limited parameters~\cite{somvanshi2024survey}.
Our scKAN model is third-order differentiable, outlines a meaningful Waddington landscape~\cite{waddington1959evolutionary} from the learned geometry.
Subsequently, we employ an explainable AI (XAI) method based on the gradients of the learned geometry to reconstruct the directed GRNs with regulation types.
By evaluating the model gradient at different cells, we can group cells by subtype or lineage, extracting a local view of the GRN for each grouping.
Our model achieves attractive results, especially in precision, on the BEELINE benchmark~\cite{pratapa2020benchmarking}.
All codes and reproducibility instructions are available at \url{https://github.com/1250326/scKAN}.

\smallskip
% \newpage
\noindent{\bf Contributions.}
Our contributions in this work are summarized as follows:
% \begin{itemize}[topsep=0.5pt, wide, leftmargin=*]
\begin{itemize}
    \item We generalize the tree-based GRN inference models with a differentiable KAN to match the underlying smooth biochemical processes.
    \item Our model infers activation and inhibition regulations with the XAI technique for arbitrary groups of cells and the whole population.
    \item We conducted more than 2,300 experiments using the BEELINE benchmark, showing scKAN surpasses the second-best signed GRN inference models from 5.40\% to 28.37\% in AUROC and from 1.97\% to 40.45\% in AUPRC.
    \item We further demonstrated the scalability of scKAN on two networks consisting of 71 and 104 nodes, respectively. 
    \item We analyze the weakness of GENIE3 and GRNBOOST2 in furcating GRNs and give an intuitive example and proof.
\end{itemize}

%------------------------------------------
\section{Related Work}
\label{sec:related}
%------------------------------------------
Understanding GRN is challenging due to data heterogeneity~\cite{nguyen2021comprehensive}, technical noise~\cite{jiang2022statistics}, and the complexity of biological systems~\cite{aalto2020gene}.
Various statistical models are developed for GRN inference to pursue stability and explainability.
Information theory-based models search for and screen out regulation patterns via mutual information (MI)~\cite{margolin2006aracne,faith2007clr}, information decomposition~\cite{chan2017pidc} and causal inference~\cite{qiu2020scribe,deshpande2022singe}.
Correlation-based models quantify the strength of relationships to model the GRN~\cite{kim2015ppcor,specht2017leap}.
Different tree algorithms are also used to directly model the dynamics of cell differentiation~\cite{huynh2010genie3,huynh2015jump3,moerman2019grnboost2,ma2020grnnonlinearode}.
The learned importance scores are interpreted as the regulation strengths.
Other models include regression~\cite{haury2012tigress,papili2018sincerities}, ordinary differential equation~\cite{matsumoto2017scode,aubin2020grisli}, matrix/tensor factorization~\cite{duren2018couplednmf,osorio2020sctenifoldnet}, time series modeling~\cite{sanchez2018grnvbem} and in-silico knockout~\cite{kamimoto2023celloracle}.
Despite the variety of GRN inference models, no single model is universally superior and performance varies between synthetic and real-world datasets~\cite{zhao2021comprehensive}.
Interested readers can refer to various GRN benchmarking papers~\cite{chen2018evaluating,pratapa2020benchmarking,nguyen2021comprehensive,zhao2021comprehensive}.

Recently, deep learning models have been applied to GRN inference.
Convolutional neural networks (CNN) derived from image processing are used to treat gene expression joint distributions as images~\cite{yuan2019cnnc,chen2021deepdrim,yuan2021tdl,xu2022dyndeepdrim,reagor2023delay}.
However, all listed CNN models, different from the traditional GRN models, are supervised models that require a list of known regulations as training input.
The dependency on known interactions limits the application of these models in real-world scenarios, as known interactions are incomplete and potentially misleading.
Other deep learning models, such as time series models~\cite{zhang2019ngnc,Monti2022dadnn}, variational autoencoder (VAE) models~\cite{shu2021deepsem,tong2024integrating}, and explainable AI models~\cite{keyl2023lrp}, can distill GRN structures during training without prior knowledge, and have also shown promise in GRN inference.
However, deep learning models lack scalability and interpretability and rely heavily on the quality and quantity of training data~\cite{dong2024deep}.
This urges the development of a new, scalable and explainable GRN inference model that can be trained on limited samples yet provide accurate and precise results.

Among existing approaches, we are particularly interested in tree-based models, such as GENIE3~\cite{huynh2010genie3} and GRNBOOST2~\cite{moerman2019grnboost2}.
GENIE3 is the pioneer that treats scRNA-seq data as tabular data, discards temporal information, and models gene expression levels as functions of other genes.
GENIE3 was the best performer in the DREAM4 multifactorial challenge~\cite{schaffter2011genenetweaver} and demonstrated competitiveness in the BEELINE benchmark~\cite{pratapa2020benchmarking}.
GENIE3 uses random forest as the predictive function and the learned importance scores are interpreted as the regulation strengths.
This method converts the unsupervised GRN inference problem into a supervised regression problem, leveraging the power of machine learning algorithms while maintaining scalability and interpretability.
GENIE3 methodology was extended to various models, such as Jump3~\cite{huynh2015jump3}, dynGENIE3~\cite{huynh2018dyngenie3}, and GRNBOOST2.
GRNBOOST2 is a similar model to GENIE3, but it uses gradient boosting as the predictive function.
Jump3 and dynGENIE3 model the change in expression levels for each cell over time, but this change is inaccessible during the destructive scRNA-seq assay.

The advantages of tree-based models are their scalability and explainability.
Since tree-based models treat each gene as a separate model, they can be easily parallelized and scaled to thousands of genes.
Using tree methods for prediction makes the model explainable, but this is also a limitation as they attempt to use piecewise continuous functions to model the smooth dynamics of cell differentiation.
In addition, since importance scores are calculated based on information gain, the score must be positive, and the activation and inhibition of the regulations cannot be distinguished.
Moreover, information gain is a statistical summary that ignores the effect of cell subtypes, lineages, and stages.
This motivates us to develop a new GRN inference model that follows the ideas of tree-based models, maintaining scalability and explainability, but can capture the continuous cellular dynamics and distinguish the activation and inhibition of the regulations.

%------------------------------------------
\section{Preliminaries}
\label{sec:pre}
%------------------------------------------
%------------------------------------------
\subsection{Notations and Problem Definition}
\label{ssec:notations}
%------------------------------------------
For an scRNA-seq dataset, we denote the number of genes and cells as $g$ and $c$, respectively.
The gene expression matrix is denoted as $X\in\mathbb{R}^{g\times c}$, where $x_p\in\mathbb{R}^g$ and $X_i\in\mathbb{R}^c$ are the $p$-th column and $i$-th row vectors of $X$, respectively.
$x_p$ can be perceived as a point of cell state in the gene space $V\subset\mathbb{R}^g$.
The gene space and the expression matrix after removing the gene $i$ are represented by $V_{\backslash i}\subset\mathbb{R}^{g-1}, X_{\backslash i}\in\mathbb{R}^{g-1\times c}$, respectively.
We denote $x_{\backslash i,p}=(\dots,X_{i-1,p},X_{i+1,p},\dots,)^T$ as the expression of cell $p$ with the gene $i$ removed.

A GRN can be represented by a weighted directed graph $\mathcal{G}=(\mathcal{V}, \mathcal{E}, w)$, where $\mathcal{V}$ is the set of $g$ nodes (genes), $\mathcal{E}\subset\mathcal{V}\times\mathcal{V}$ is the set of directed edges (regulatory relationship), and $w:\mathcal{E}\rightarrow \mathbb{R} \setminus \{0\}$ is the weight function (regulation strength).
The adjacency matrix $A\in\mathbb{R}^{g\times g}$ is a matrix representation of the regulatory graph $\mathcal{G}$, where $A_{i,j}=w(i,j)$ if $(i,j)\in\mathcal{E}$ and $A_{i,j}=0$ otherwise.

We formulate our research problem as finding the adjacency matrix that best recovers the underlying GRN from the observed scRNA-seq gene expression data without prior knowledge of the network structure.

%------------------------------------------
\subsection{Formulation of One-vs-Rest Models}
\label{ssec:formulation_tree_model}
%------------------------------------------
All tree-based models, such as GENIE3, Jump3, dynGENIE3 and GRNBOOST2 can be formulated as regression problems summarized as:
\begin{equation}
    \label{eq:ovr_model}
        \hat{X}_{i,p} = f_i(x_{\backslash i,p})
\end{equation}
where $i\in\{1,\dots,g\},\ p\in\{1,\dots,c\}$, and $f_i:V_{\backslash i}\rightarrow\mathbb{R}$ is a tree model.
$\hat{X}_{i,p}$ is the expression level for GENIE3 and GRNBOOST2, and the change in expression level for Jump3 and dynGENIE3.
The learned importance score for gene $j$ in model $i$ is denoted as $\mathcal{I}(f_i,j)$, and is interpreted as the strength of the regulation of gene $j$ on gene $i$.
We collectively refer to models formulated in the form of Eq.~(\ref{eq:ovr_model}) as One-vs-Rest (OvR) models in the following sections.
Our model is built on the formulation of the OvR model, and we replace the tree model $f_i$ with a Kolmogorov-Arnold network model.

%------------------------------------------
\subsection{Kolmogorov-Arnold Networks}
\label{ssec:kan}
%------------------------------------------
Kolmogorov-Arnold network (KAN)~\cite{liu2024kan} is a recent advance in neural network research, potentially offering a new alternative to traditional multi-layer perceptrons (MLP).
Both of them have a similar theoretical foundation.
According to the Universal Approximation Theorem~\cite{hornik1989multilayer} and the Kolmogorov-Arnold representation theorem~\cite{kolmogorov1961representation}, feed-forward neural networks and KANs can both approximate any continuous function under mild constraints.
There are extensive debates on the performance of KAN compared to MLP across different tasks.
Multiple works show that KAN generally excels in symbolic regression~\cite{yu2024kancomparison} and graph regression tasks~\cite{bresson2024kagnns} and performs better with limited parameters~\cite{shukla2024comprehensive}.
Compared with the usual MLP models, KAN models are also more expressive with the same number of layers and offer higher convergent rates~\cite{wang2024expressiveness}.

A MLP layer $f_{\text{MLP}_\ell}$ with $n_i$ inputs and $n_{i+1}$ outputs consists of a weight matrix $W_\ell\in\mathbb{R}^{n_{i+1}\times n_i}$, a bias vector $b_\ell\in\mathbb{R}^{n_{i+1}}$ and an activation function $\sigma_\ell:\mathbb{R}^{n_{i+1}}\rightarrow\mathbb{R}^{n_{i+1}}$:
\begin{equation}
    \label{eq:mlp_layer}
    f_{\text{MLP}_\ell}(x)=\sigma_\ell(W_\ell\,x+b_\ell),
\end{equation}
where input features are aggregated by linear combinations with nonlinear activations.
A MLP model $f_{\text{MLP}}$ is a composition of $L$ MLP layers:
\begin{equation}
    \label{eq:mlp}
    f_{\text{MLP}}(x)=f_{\text{MLP}_L}\circ f_{\text{MLP}_{L-1}}\circ\dots\circ f_{\text{MLP}_1}(x).
\end{equation}

Unlike MLP, KAN defines activation functions $\phi:\mathbb{R}\rightarrow\mathbb{R}$ to aggregate input features, 
which contains $k$-th order B-splines $B_i$ with grid size $G$, a SiLU nonlinearity and scalar weights $w_b,w_i$:
\begin{equation}
    \label{eq:kan_spline}
    \phi(x)=w_b\,\text{SiLU}(x)+\sum_{i=0}^{G+k-1} w_i B_i(x).
\end{equation}
A KAN layer $\Phi_\ell:\mathbb{R}^{n_i}\rightarrow\mathbb{R}^{n_{i+1}}$ with $n_i$ inputs and $n_{i+1}$ outputs is an operator consisting of a matrix of activation functions:
\begin{equation}
    \label{eq:kan_layer}
    \Phi_\ell= \begin{bmatrix}
        \phi_{\ell,1,1} & \dots & \phi_{\ell,1,n_i} \\
        \vdots & \ddots & \vdots \\
        \phi_{\ell,n_{i+1},1} & \dots & \phi_{\ell,n_{i+1},n_i}
    \end{bmatrix},\quad \phi_{\ell,q,p}:\mathbb{R}\rightarrow\mathbb{R}.
\end{equation}
The KAN model $f_{\text{KAN}}:\mathbb{R}^n\rightarrow\mathbb{R}$ with input $x$ is a composition of $L$ KAN layers:
\begin{equation}
    \label{eq:kan_model}
    f_{\text{KAN}}(x)=\Phi_L\circ\Phi_{L-1}\circ\dots\circ\Phi_1(x).
\end{equation}

%------------------------------------------
\section{Methodology}
\label{sec:method}
%------------------------------------------
\begin{figure*}[!t]
    \centering
    \includegraphics[width=0.80\textwidth]{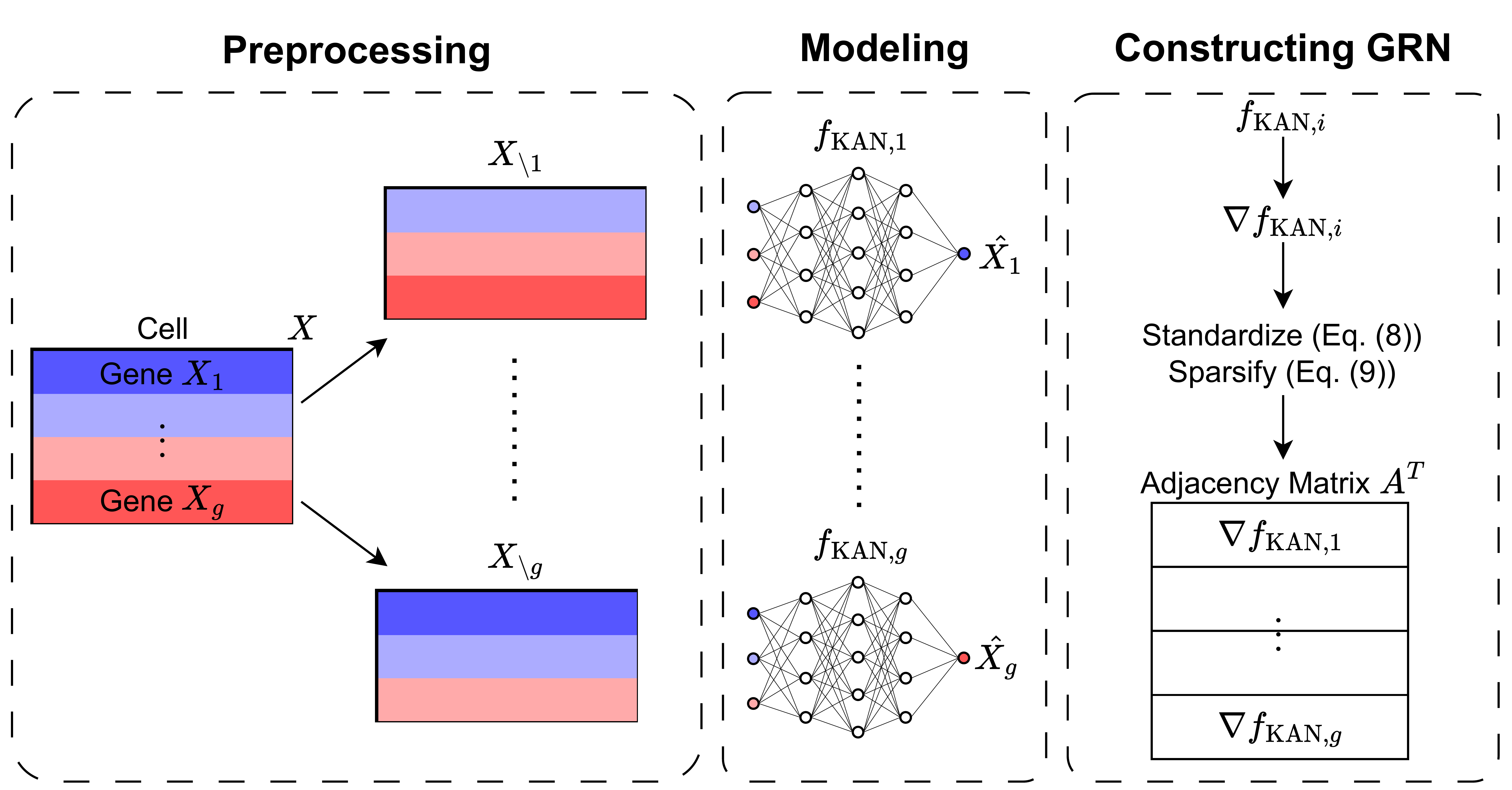}
    \caption{A graphical summary of scKAN.
             For each gene $i$, we remove its gene expression and train an independent model $f_{\text{KAN},i}$ to recover its expression.
             We extract the gradients from the models to infer the GRN after standardization and sparsification.
             }
    \label{fig:model}
\end{figure*}

Our method consists of the following three steps.

%\subsection{Preprocessing}
\smallskip
\noindent{\bf Preprocessing.}
We start by creating a copy of the gene expression matrix for each gene $i$, and removing the $i$-th gene from the matrix.
This design ensures scalability by parallelizing the training process.

%\subsection{Modeling}
\smallskip
\noindent{\bf Modeling.}
We follow the formulation of OvR models in Eq.~(\ref{eq:ovr_model}) to build individual models for each gene.
We replace the predictor $f_i$ with a KAN model, which can approximate any continuous function under mild constraints.
Hence, assuming that such a function $f_i$ exists, our replacement of the neural network can also approximate $f_i$ effectively using the Kolmogorov-Arnold representation theorem.
The formulation of scKAN is given as follows:
\begin{equation}
    \label{eq:scKAN}
    \hat{X}_{i,p} = f_{\text{KAN},i}(x_{\backslash i,p}),
\end{equation}
where $i\in\{1,\dots,g\},\ p\in\{1,\dots,c\}$, and $f_{\text{KAN},i}:V_{\backslash i}\rightarrow\mathbb{R}$.

We set the spline order as 3 and grid size as 10 for each $\phi_{\ell,q,p}$.
The former ensures the continuity of the gradients and third-order differentiability of the KAN predictor $f_{\text{KAN},i}$, while the latter balances the model granularity.

ScRNA-seq datasets are often high-dimensional with limited samples.
An expressive MLP model often requires a deep structure with a large number of parameters~\cite{wang2024expressiveness}, which is prone to overfitting.
Thus, the shallow KAN model is a better choice for modeling small datasets while assisting explainability.
To limit the number of parameters in scKAN, we use only 3 hidden layers.
Following Liu et al.'s recommendation~\cite{liu2024kan}, the number of neurons in each hidden layer is set to $2d+1$, $2(2d+1)+1$, and $2d+1$, respectively, where $d=g-1$.
In this configuration, a single KAN predictor in a 7-gene data set only requires 75 seconds of training time.
The model run times for different numbers of genes can be found in Fig.~\ref{fig:time}.

%\subsection{Constructing GRN}
\smallskip
\noindent{\bf Constructing GRN.}
We quantify the importance scores of the trained predictors based on the gradient $\nabla f_{\text{KAN},i}$, which can be interpreted as the slope on the Waddington landscape.
The gradient evaluated at each data point $x_{\backslash i,p}\in V_{\backslash i}$ can be easily retrieved using automatic differentiation.
The positive (negative) gradient of $f_{\text{KAN},i}$ with respect to gene $j$ indicates the activation (inhibition) of gene $j$ on gene $i$, while the magnitude of the gradient indicates the regulation strength.
However, the signals hidden under the gradients are sensitive to the data noise and overfitting.
Assuming a sparse GRN, the targeted gene $i$ is regulated by a few genes, and their magnitudes of the gradients are significantly larger than those of the others.
Thus, we detect signals from the gradients' magnitude across genes via \textit{standardization}:
\begin{equation}
    \label{eq:standardization}
    s_{j,i,p} = \frac{|\nabla_j f_{\text{KAN},i}(x_{\backslash i,p})| - \text{mean}(|\nabla f_{\text{KAN},i}(x_{\backslash i,p})|)}{\text{std}(|\nabla f_{\text{KAN},i}(x_{\backslash i,p})|)},
\end{equation}
for all $p\in\{1,\dots,c\}$, where $s_{j,i,p}$ is the z-score of the gradient magnitude, $\nabla_j f_{\text{KAN},i}$ is the gradient of $f_{\text{KAN},i}$ with respect to gene $j$, and $\text{mean}(\cdot)$ and $\text{std}(\cdot)$ are the mean and standard deviation functions on genes, respectively.

Since $s_{j,i,p}$ is generally nonzero, simply averaging the gradients across cells will give a fully connected GRN.
To sparsify the GRN, we count the portion of cells where the standard score $s_{j,i,p}$ is greater than 1.
Since the sign indicates the type of regulation, we use majority voting to determine the type of regulation.
The adjacency matrix $A$ is then constructed by \textit{sparsification}:
\begin{equation}
    \label{eq:sparsification}
    A_{j,i} = \text{mode}(\text{sign}(\nabla_j f_{\text{KAN},i}(x_{\backslash i,p}))) \cdot \frac{1}{c} \cdot \sum_p{1_{s_{i,j,p} > 1}},
\end{equation}
where $i,j\in\{1,\dots,g\}$.

Under this construction, the diagonal elements of the adjacency matrix are zero because gene $i$ is not a variable of $f_{\text{KAN},i}$.
In addition, the adjacency matrix is asymmetric because the entries $A_{i,j}$ and $A_{j,i}$ are calculated based on the gradients of $f_{\text{KAN},j}$ and $f_{\text{KAN},i}$, respectively.
The illustration of scKAN is shown in Fig.~\ref{fig:model} and the outline of the algorithm can be found in Alg.~\ref{alg:scKAN} in Appendix~\ref{sec:appendix}.

%------------------------------------------
\section{Experiments}
\label{sec:exp}
%------------------------------------------
We evaluated scKAN on the BEELINE benchmark~\cite{pratapa2020benchmarking}, which is the de facto benchmark for GRN inference models based on scRNA-seq data without prior knowledge~\cite{fan2021gene,shu2022boosting}.
It contains 6 types of synthetic networks, namely linear (LI, $g=7$), linear long (LL, $g=18$), cyclic (CY, $g=6$), bifurcating (BF, $g=7$), bifurcating converging (BFC, $g=10$), and trifurcating (TF, $g=8$), which capture different cell differentiation processes.
It also contains 4 types of curated networks, namely mCAD ($g=5$), VSC ($g=8$), HSC ($g=11$), and GSD ($g=19$), extracted from reported biological processes.
For each network type, BEELINE used BoolODE~\cite{pratapa2020benchmarking} to generate simulated scRNA-seq datasets.
BoolODE is a popular scRNA-seq simulator that generates synthetic datasets with given Boolean functions describing the regulation mechanisms.
Then, it applies the Hill equation~\cite{hill1910possible} and the chemical Langevin equation~\cite{langevin1908theorie} to form a system of ODEs and simulate the gene expression data.
In this study, we used 10 datasets with $c=2000$ cells for each type of network.

We compare our model with the GRN baselines on the BEELINE benchmark, including PPCOR~\cite{kim2015ppcor}, LEAP~\cite{specht2017leap}, SCODE~\cite{matsumoto2017scode}, GRISLI~\cite{aubin2020grisli}, GRNVBEM~\cite{sanchez2018grnvbem}, SINCERITIES~\cite{papili2018sincerities}, SINGE~\cite{deshpande2022singe}, PIDC~\cite{chan2017pidc}, Scribe~\cite{qiu2020scribe}, GENIE3~\cite{huynh2010genie3}, and GRNBOOST2~\cite{moerman2019grnboost2}.
Among all listed baselines, only SCODE, GRNVBEM, SINCERITIES and our scKAN can further determine the sign of edges (activation and inhibition).
Here, we excluded SCNS~\cite{woodhouse2018scns} as it requires prior knowledge of GRN, which is not consistent with our research problem.

%------------------------------------------
\subsection{Experiment Setup}
\label{ssec:preprocessing}
%------------------------------------------
In the following experiments, the random train-test split ratio is set to $8:2$.
Models were trained with 3,000 epochs using the Adam optimizer with a $10^{-4}$ learning rate and gradient clipping at $100$.
Mean squared error was used as the loss function.
Early stopping was applied when the gap between train and test loss exceeded $0.0005$ for 10 epochs.
Although test loss was used in the decision of early stopping, this does not cause label leakage because test loss is used to detect overfitting and quantify the quality of fit on gene expression levels, and the ground-truth GRN used for evaluation is still unseen by the model.
Our models were trained on a single NVIDIA V100 SXM2 GPU with 16GB memory, while all baseline models were run on a 4.90 GHz 20 threads Intel i7-12700 CPU with 64GB memory.
We trained and tested our model on 10 different datasets generated by BoolODE.
For each cell $p$, we calculate the gradient of the learned predictor and sparsify the adjacency matrix.
The entries in the adjacency matrix are calculated following Eqs.~(\ref{eq:standardization},\ref{eq:sparsification}).

To align our research with the BEELINE benchmark, we used the same datasets and optimized hyperparameter sets in all baseline models.

We evaluated all models using the AUROC and AUPRC following the BEELINE setting, assisted by the Structured Hamming Distance (SHD) and False Discovery Rate (FDR).
For a model output $A\in\mathbb{R}^{g\times g}$ and a ground-truth network $G\in[0,1]^{g\times g}$, we can count the true positive $TP(t)$, true negative $TN(t)$, false positive $FP(t)$ and false negative $FN(t)$ edges under different thresholds $t$ in $A$.
AUROC is defined by the area under the curve of true positive rate ${\it TPR}(t)=\frac{TP(t)}{TP(t)+FN(t)}$ against false positive rate ${\it FPR}(t)=\frac{FP(t)}{FP(t)+TN(t)}$;
Similarly, AUPRC is defined by the area under the curve of precision $P(t)=\frac{TP(t)}{TP(t)+FP(t)}$ against recall $R(t)=\frac{TP(t)}{TP(t)+FN(t)}$:
\begin{equation}
    \label{eq:metric}
   {\it  AUROC} = \int_0^1 TPR(FPR)\ dFPR, \quad
   {\it AUPRC} = \int_0^1 P(R)\ dR,
\end{equation}

AUROC measures the model's accuracy, while AUPRC measures the prediction precision.
Both range from 0 to 100\%, with higher values indicating better performance.
A random model has an AUROC of 50\%, and an AUPRC equal to the network density.
SHD measures the closeness of the predicted and ground-truth networks, while FDR measures the edge prediction error.
SHD is a positive real number and the FDR ranges from 0 to 100\%.
Smaller values indicate better results in both metrics.

This paper employs multi-class classification metrics, such as AUROC and AUPRC, for directed and signed ground-truths, while SHD and FDR are utilized for directed ground-truths.

%------------------------------------------
\subsection{Results}
\label{ssec:results}
%------------------------------------------

\begin{figure}[!t]
    \centering
    \includegraphics[width=\textwidth]{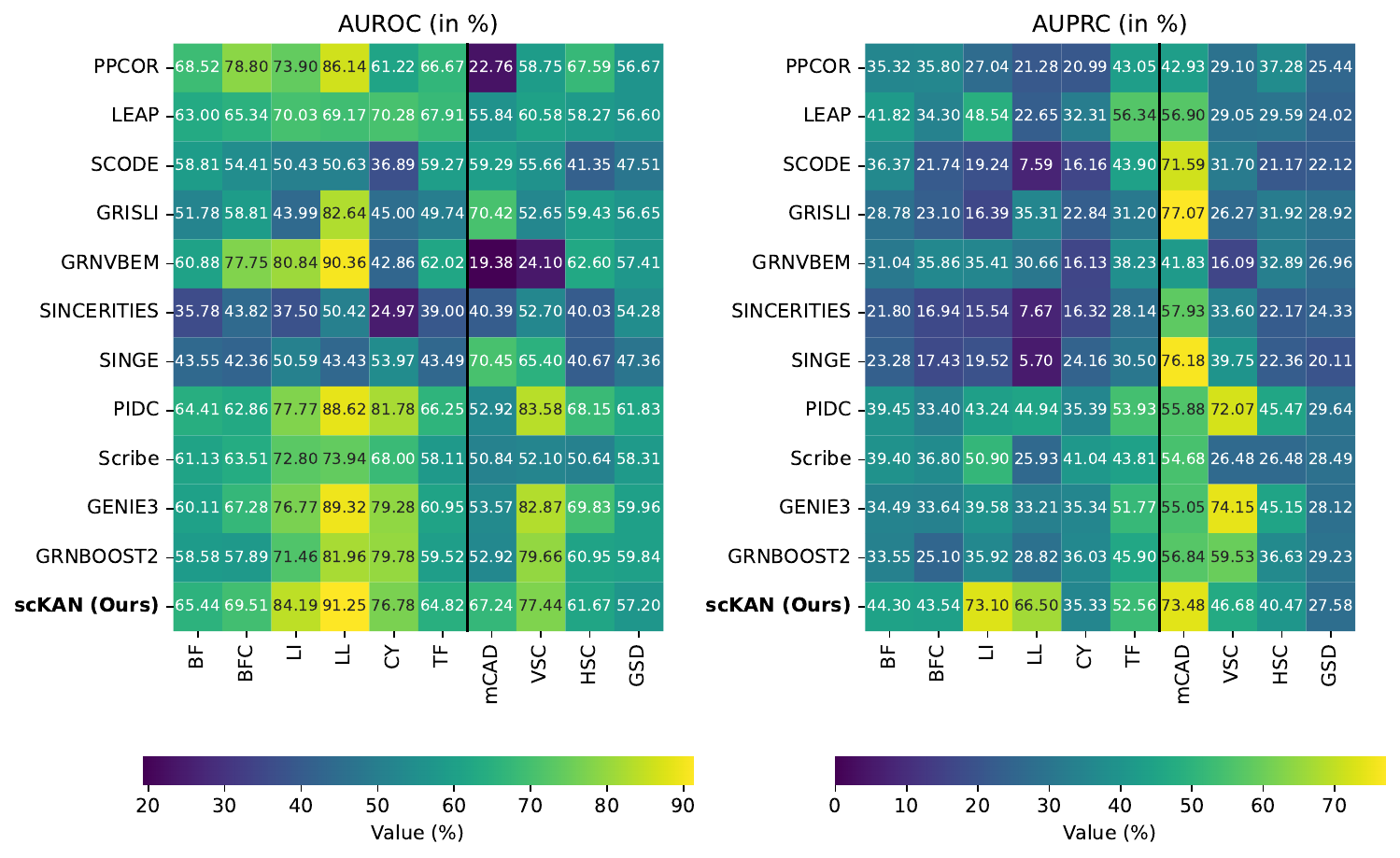}
    \caption{Average AUROC (left) and AUPRC (right) of directed GRN inference models with unsigned ground-truth on different datasets out of 10 runs.
             Synthetic and curated datasets are separated by a vertical line.}
    \label{fig:directed}
\end{figure}

\noindent{\bf Directed unsigned GRNs.}
The performance comparison of the directed GRN inference models with unsigned ground-truth on the BEELINE benchmark is shown in Fig.~\ref{fig:directed}.
Comparisons on SHD and FDR can be found in Figs.~\ref{fig:directed_shd},~\ref{fig:directed_fdr} in Appendix~\ref{sec:appendix}.
We observed that scKAN outperforms all other models in terms of AUROC and AUPRC in most of the synthetic datasets, and scKAN has the best average AUROC and AUPRC in all datasets.
Notably, our model has much higher AUPRC values in linear (LI) and linear long (LL) datasets, surpassing the second-best model by 22.2\% and 21.56\%, respectively, suggesting that scKAN is strong in identifying long-range dependencies.
In curated datasets, scKAN also reaches top-tier performance in AUROC and AUPRC.

\begin{figure}[!t]
    \centering
    \includegraphics[width=\textwidth]{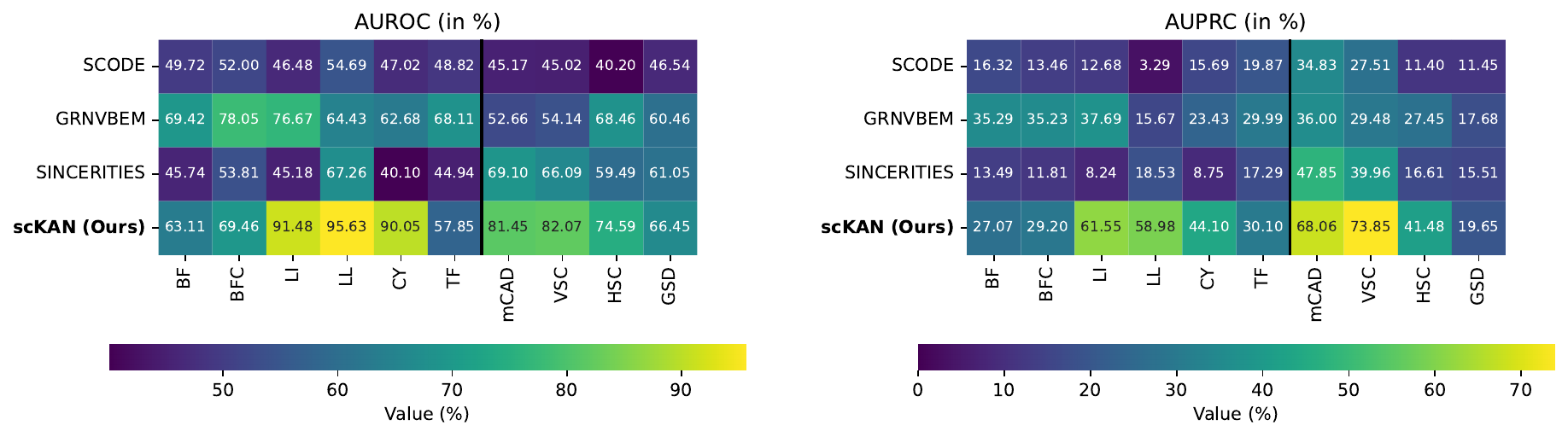}
    \caption{Average AUROC (left) and AUPRC (right) of directed GRN inference models with signed ground-truth on different datasets out of 10 runs.
             }
    \label{fig:signed_directed}
\end{figure}

\smallskip\noindent{\bf Directed signed GRNs.}
The performance comparison of the directed GRN inference models with signed ground-truth on the BEELINE benchmark is shown in Fig.~\ref{fig:signed_directed}.
We notice that SCODE and SINCERITIES generally have poor precision in most datasets.
In contrast, scKAN demonstrates the highest accuracy and precision in LI, LL, CY, and all curated datasets, surpassing the second-best model from 5.40\% to 28.37\% in AUROC and from 1.97\% to 40.45\% in AUPRC.
However, scKAN faced difficulties in furcating (BF, BFC, TF) datasets.
In Section~\ref{ssec:ovr_weakness}, we will justify that all OvR models are weak in furcating datasets with concrete examples.

The standard deviation of the AUROC and AUPRC values for directed GRN inference models with unsigned and signed ground-truths can be found in Figs.~\ref{fig:directed_std},~\ref{fig:signed_directed_std} in Appendix~\ref{sec:appendix}. 

\smallskip\noindent{\bf Scalability.}
We extracted the consensus GRN from the CollecTRI~\cite{muller2023expanding} database and retained two communities with 71 and 104 genes.
We used BoolODE to generate 10 datasets for each community with 2,000 cells and compared scKAN with the baselines in these datasets.
GRISLI and SINGE hit memory and time constraints and are excluded from the comparison.

The average and standard deviation of the AUROC and AUPRC are shown in Tables~\ref{tab:scalability_unsigned},~\ref{tab:scalability_signed} in Appendix~A. 
We observed that SCODE, GRNVBEM, SINCERITIES, and Scribe are not scalable to large datasets, as their AUROC and AUPRC values are close to the random model.
The performance of PPCOR, LEAP, and PIDC significantly drops when scaling to 104 genes, while all OvR models perform similarly in two large datasets.
This validates the scalability of OvR models in large datasets.

GENIE3 is the best performer of both metrics in the two datasets, but it cannot infer edge types.
scKAN is the only model that can infer edge types (signs) among OvR models, and is the most scalable model among all signed models with an AUROC of 63.18\% and AUPRC of 6.18\% on average.
This again showcases the scalability and accuracy of the family of OvR models.

\begin{figure}[!t]
    \centering
    \includegraphics[width=0.6\textwidth]{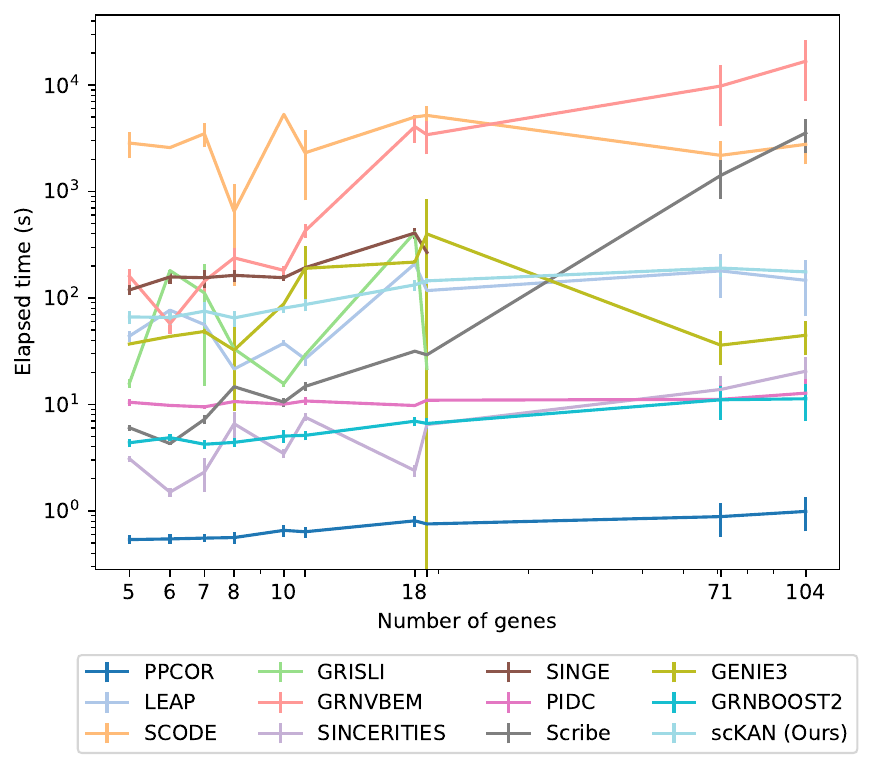}
    \caption{Model elapsed time on different dataset sizes in log-log scale. Vertical bars indicate standard deviations at each data point.}
    \label{fig:time}
\end{figure}

We further summarize the model run times in Fig.~\ref{fig:time}.
Because of each model's stochastic nature, model run times vary significantly for the same number of genes and do not strictly increase as the number of genes increases.
Our model has a run-time of 176 seconds for datasets with 104 genes, similar to LEAP, and is trendwise more scalable than GRNVBEM, SINCERITIES, and Scribe.

%------------------------------------------
\subsection{Ablation Study}
\label{ssec:ablation}
%------------------------------------------
\begin{figure}[!t]
    \centering
    \includegraphics[width=\textwidth]{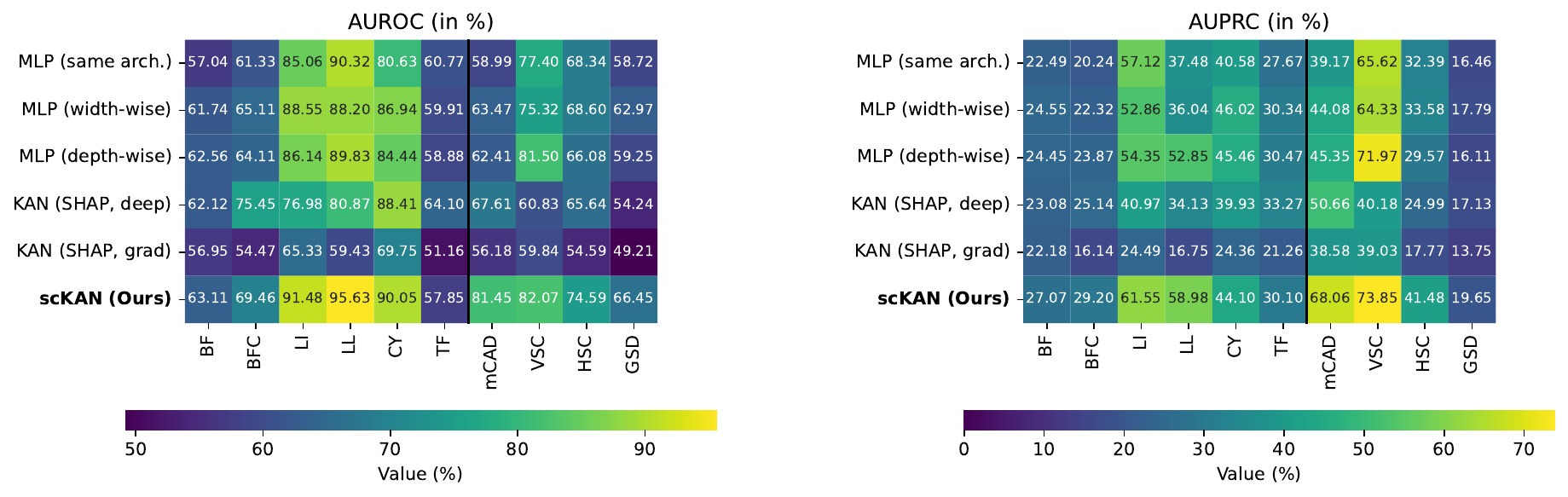}
    \caption{Average AUROC (left) and AUPRC (right) of models in the ablation study on different datasets out of 10 runs.
             }
    \label{fig:signed_directed_ablation}
\end{figure}

Compared with other OvR models, our model replaces the tree model with a KAN model, an alternative to MLP, and replaces the importance score with the predictor's gradient.
We conducted an ablation study to prove the necessity of these modifications.
Model performances under different configurations of inferring directed and signed ground-truth networks are shown in Fig.~\ref{fig:signed_directed_ablation}.
The standard deviation of the AUROC and AUPRC values can be found in Fig.~\ref{fig:signed_directed_ablation_std} in Appendix~\ref{sec:appendix}. 

\smallskip
\noindent{\bf KAN vs. MLP.}
We first compare a KAN predictor with an MLP predictor with the same architecture, labeled `MLP (same arch.)'.
For a fair comparison, we also compare the KAN predictor with an MLP predictor with a similar number of parameters by adjusting the layer width and depth, labeled as `MLP (width-wise)' and `MLP (depth-wise)', respectively in Fig.~\ref{fig:signed_directed_ablation}.
We observed a consistent weakness in furcating datasets for all compared models in line with the claim in Section~\ref{ssec:results}, and nearly all MLP models are slightly inferior to the KAN model in AUROC and AUPRC.
In addition, the KAN model is more promising in curated datasets, suggesting that the KAN model can better capture complex cellular dynamics.
This observation aligns with Wang et al.~\cite{wang2024expressiveness}'s conclusion that KAN is more expressive than MLP under the same number of layers.

\smallskip
\noindent{\bf Gradient vs. explainable AI.}
Various XAI tools have been developed to account for the contribution of each feature to the model prediction.
Gradients are fundamental in many XAI tools~\cite{sundararajan2017axiomatic,selvaraju2017grad,kim2018interpretability,mudrakarta2018did}.
In contrast, Shapley additive explanations (SHAP)~\cite{Lundberg2017shap} is another popular model-agnostic tool that explains the importance of each feature in model prediction.
We replaced the calculated gradient with SHAP values using the deep explainer and gradient explainer in the SHAP library.
Their performances are labeled `KAN (SHAP, deep)' and `KAN (SHAP, grad)' in Fig.~\ref{fig:signed_directed_ablation}.

Compared with scKAN explained by raw gradient values, although the deep explainer is more accurate in furcating datasets, it is less precise and significantly inferior in other datasets.
This suggests that the predictor gradient is a better choice than the SHAP values.

%------------------------------------------
\section{Discussion and Limitations}
\label{sec:discussion}
%------------------------------------------
%------------------------------------------
\subsection{GRNs for Rare Cell Subtypes}
\label{ssec:rare_subtype}
%------------------------------------------
\begin{figure}[!t]
    \centering
    \includegraphics[width=\textwidth]{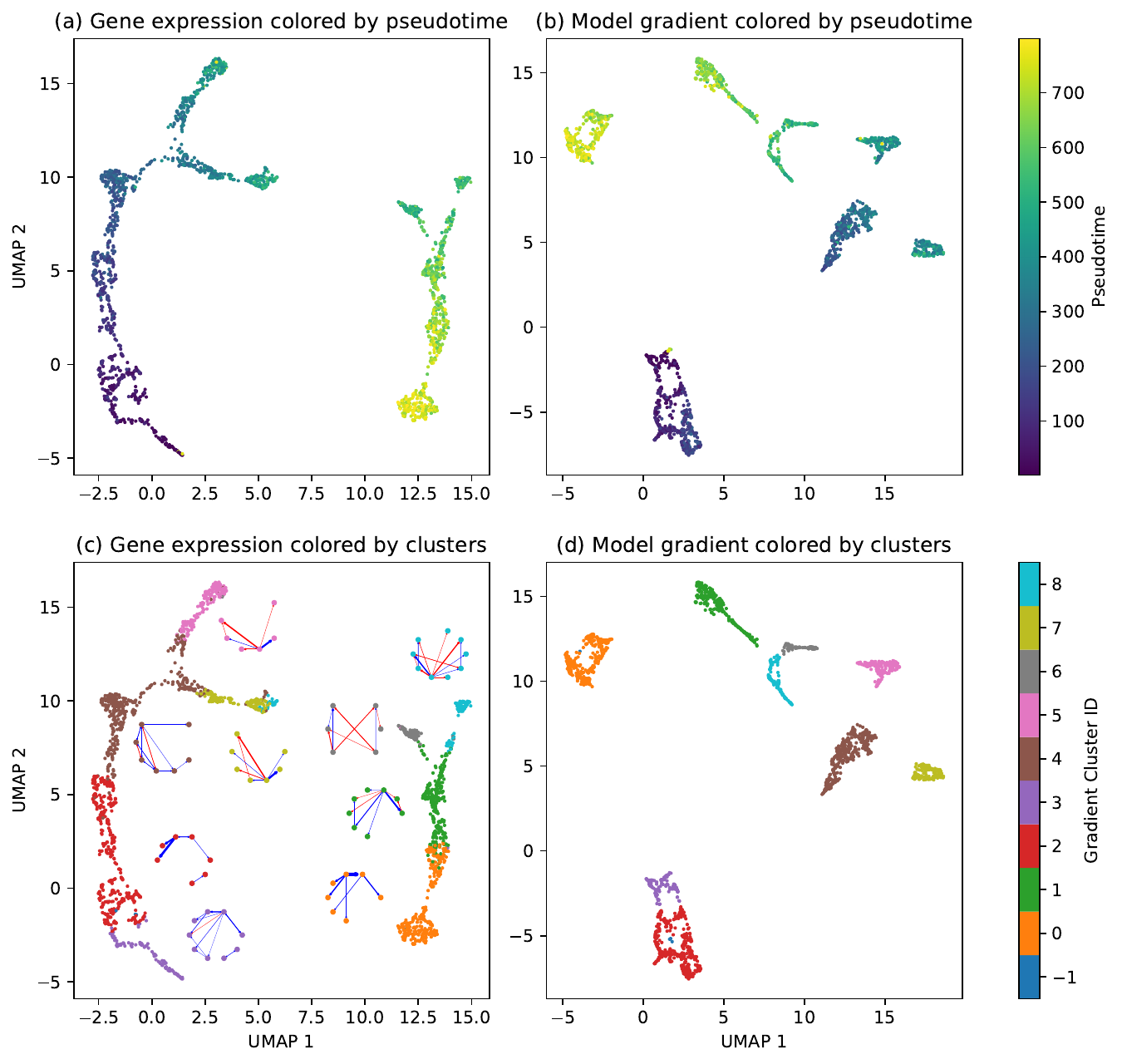}
    \caption{UMAP visualization of the gene expression and gradient pattern colored by pseudotime and gradient cluster.
             Inferred cell-type specific GRNs are shown in (c).}
    \label{fig:cell_type_specific_GRN}
\end{figure}

Identifying GRNs in rare cell subtypes is highly valuable for pathology, as it allows mechanistic understanding, diagnostic refinement, and therapeutic target discovery for disease-driving cell populations.
However, the low abundance of these subtypes in scRNA-seq data makes it challenging to identify their GRNs.
GRNs inferred by GENIE3 and GRNBOOST2, in particular, are often dominated by the majority cell types, leading to a loss of information on rare subtypes.
Conversely, we can evaluate the gradient at any cell sample from the learned geometry of our scKAN model and obtain the GRN at single-cell resolution.
Since our model gives a smooth geometry of the gene expression landscape, the gradients preserve both global and local information about the dynamics.

We visualized gene expression and gradient patterns in the bifurcating converging (BFC) dataset using UMAP~\cite{mcinnes2018umap} in Fig.~\ref{fig:cell_type_specific_GRN}.
We clustered the gradients with DBSCAN~\cite{ester1996density}, which does not require prior knowledge of the number of clusters and can robustly identify outliers.
In Fig.~\ref{fig:cell_type_specific_GRN}(d), we clustered the gradients into 9 clusters, and the outliers are isolated in group `-1'.
We colored gene expression by clusters in Fig.~\ref{fig:cell_type_specific_GRN}(c), and it shows that gradient clusters can identify different cell stages.
For each cluster, we inferred a cell-type-specific GRN and visualized the GRN in Fig.~\ref{fig:cell_type_specific_GRN}(c).
Cell-type-specific GRNs are significantly different in each cluster and identify key drivers for bifurcating cell fates.
This technique can be applied to the analysis of rare cell subtypes, providing a new perspective on the mechanism of these cell subtypes.

%------------------------------------------
\subsection{Weakness of OvR Models}
\label{ssec:ovr_weakness}
%------------------------------------------
\begin{figure}[!h]
    \centering
    \subfigure[]{\includegraphics[width=0.48\textwidth]{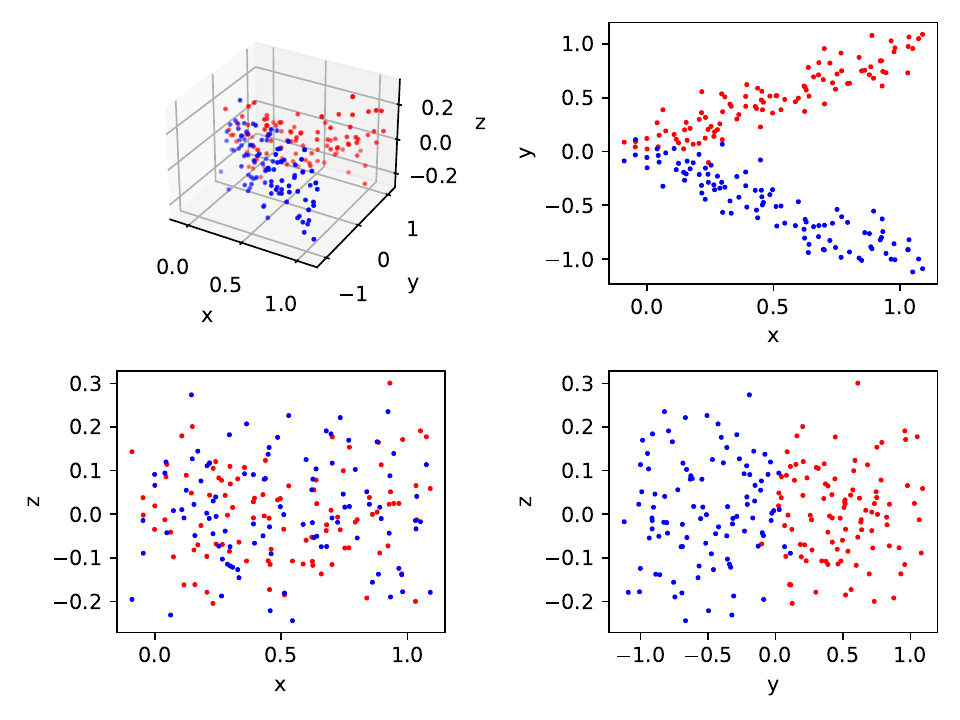}} 
    \subfigure[]{
        \raisebox{0.5\height}{
        \begin{tikzpicture}[node distance=1cm, >=stealth]
        % Nodes
        \node[circle, draw, minimum size=1cm] (x) at (0, 0) {$x$};
        \node[circle, draw, minimum size=1cm] (y) at (2, 0) {$y$};
        \node[circle, draw, minimum size=1cm] (z) at (1, 1.5) {$z$};
        % Arrows
        \draw[->, bend left] (x) to (y);
        \draw[->, bend left] (y) to (x);
        \end{tikzpicture}
        }
    }
    \caption{(a) Simulations of gene expression with blue and red branches. (b) Underlying GRN of the 3-gene toy example.}
    \label{fig:branches_and_3genes}
\end{figure}
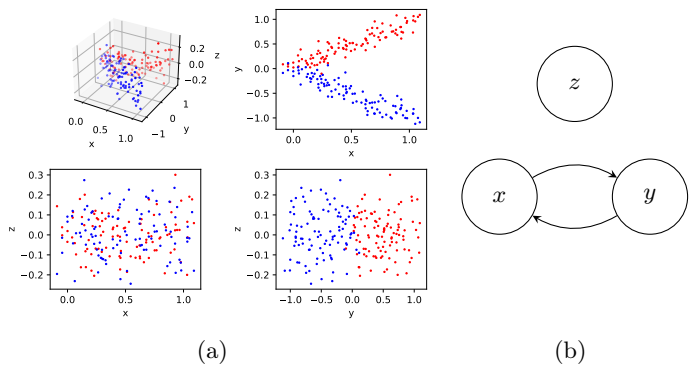

We empirically demonstrated the capability and scalability of scKAN to infer directed GRNs with activation and inhibition edges accurately and precisely.
Nevertheless, we observed a consistent weakness of GENIE3, GRNBOOST2 and scKAN in furcating datasets.
In the following, we will justify that such weakness is due to the nature of the OvR model.

Consider the following 3-gene toy example with cells on red  $(x,y_r,z_r)$ and blue $(x,y_b,z_b)$ branches visualized in Fig.~\ref{fig:branches_and_3genes}(a):
\begin{equation}
    y_r = x + \epsilon_r, \qquad y_b = -x + \epsilon_b,
\end{equation}
where $x\in [0,1]$ and $\epsilon_r,\epsilon_b,z_r,z_b\sim \mathcal{N}(0,0.1)$ are normal distributed.
Genes $x$ and $y$ are interrelated while gene $z$ is isolated, so both branches have the same GRN shown in Fig.~\ref{fig:branches_and_3genes}(b).

For the dataset with both branches and the dataset with the red branch, GENIE3 infers ${\it GRN}_{both},{\it GRN}_{r}$, respectively:
\begin{equation*}
    {\it GRN}_{both}=\begin{bmatrix}
0 & 0.48 & 0.47 \\
0.92 & 0 & 0.53 \\
0.08 & 0.52 & 0 
\end{bmatrix},\ 
    {\it GRN}_{r} = \begin{bmatrix}
0 & 0.92 & 0.58 \\
0.92 & 0 & 0.42 \\
0.08 & 0.08 & 0 
\end{bmatrix}
\end{equation*}

We can observe that only ${\it GRN}_r$ is confident ($>0.9$) for the regulations $x\rightarrow y$ and $y\rightarrow x$, while ${\it GRN}_{both}$ only identifies the regulation $y\rightarrow x$.
If we only consider the red cells in Fig.~\ref{fig:branches_and_3genes}(a), cells form a line from the x-y and y-x planes, and GENIE3 can accurately capture the linear relationships.
However, when considering both red and blue cells, cells form a V-shape from the y-x plane and a `$<$'-shape from the x-y plane.
The former can be modeled by a proper function, while the latter cannot because a single value x corresponds to multiple y values.
This explains why ${\it GRN}_{both}$ cannot detect $x\rightarrow y$ regulation.

Abstractly speaking, in an OvR model, the predictor $f_i$ is trained to predict the expression of gene $g_i$ based on the expression of all other genes in the space $V_{\backslash i}$.
$V_{\backslash i}$ can be viewed as the `shadow' of the gene expression space $V$ projected on the $i$-th axis, and the OvR model fits a function to recover the height from the shadow.
In furcating datasets, the distribution of the expression data is divided into multiple branches, and the shadows of different branches overlap.
For points in this overlapping region, the gene expression of the $i$-th gene corresponds to different values at each branch, which violates the function definition. 
This leads to a consistent weakness of the OvR models in all furcating datasets.

%------------------------------------------
\subsection{Model Limitations}
\label{ssec:limitations}
%------------------------------------------
Despite the shared weakness of OvR models, we also acknowledge the following limitations.
First, we evaluated the performance of scKAN only on synthetic and curated datasets, excluding real-world datasets.
BEELINE relies on BoolODE to generate the datasets, and the expression is not zero-inflated. 
Catering to zero inflation requires special designs in our model.
In addition, scKAN, as well as other OvR models, cannot detect self-loops because the expression value of gene $i$ is excluded from the predictor $f_i$.
More importantly, OvR models ignore the temporal information such as sampling time points and pseudotime, and the inferred biological system is not identifiable.
Prudence should be taken when interpreting the inferred GRNs.

%------------------------------------------
\section{Conclusion and Future Work}
\label{sec:conclusion}
%------------------------------------------
In this work, we introduce a novel model, scKAN, to infer GRNs from scRNA-seq data without prior knowledge of the GRNs.
Our model is a generalization of the GENIE3 and GRNBOOST2 models. 
Our approach replaces the piecewise continuous tree models with continuous KAN neural networks, which align with the Waddington landscape, capture the characteristics of the intrinsic data generation process and ensure the continuity and smoothness of the gradients. 
Moreover, our model allows inferring directed GRNs with activation and inhibition edges, which is not possible for current OvR models and most state-of-the-art GRN inference models.
We evaluated our model against other baseline models on the BEELINE benchmark.
scKAN has achieved top-tier performance across various datasets, which shows its capability, robustness and scalability.
Furthermore, scKAN excels in precision and inferring long-range dependencies, which are the primary challenges in GRN inference.
On the other hand, we also noticed the weakness of OvR models in furcating datasets, and gave an intuitive example and explanations.

In the future, we will utilize higher-order derivatives of the predictors, which encode more information from the cell dynamics, to infer GRNs.
Besides, we will evaluate our model on real-world zero-inflated gene expression and compare its performance with other state-of-the-art models.

The code and datasets used in this study are publicly available at \url{https://github.com/1250326/scKAN}.

\begin{credits}
\subsubsection{\ackname} Authors Tsz Pan Tong and Jun Pang acknowledge financial support of the Institute for Advanced Studies of the University of Luxembourg through an Audacity Grant (AUDACITY-2021).

%------------------------------------------
\subsubsection{\discintname}
%------------------------------------------
The authors have no competing interests to declare that are relevant to the content of this article.
\end{credits}

\newpage
%------------------------------------------
\bibliography{paper}
\bibliographystyle{splncs04}
%------------------------------------------

\newpage
\appendix
%------------------------------------------
\section{Appendix}
\label{sec:appendix}
%------------------------------------------
\begin{algorithm}[!ht]
    \caption{scKAN: Kolmogorov-Arnold Network for GRN Inference}
    \label{alg:scKAN}
    \begin{algorithmic}[1]
        \State \textbf{Input:} scRNA-seq data $X\in\mathbb{R}^{g\times c}$
        \State \textbf{Output:} Adjacency matrix $A\in\mathbb{R}^{g\times g}$
        \For{$i=1$ to $g$}
            \State $f_i = \text{Train}(X_{\backslash i}, X_i)$
            \For{$p=1$ to $c$}
                \State $\nabla f_{\text{KAN},i}(x_{\backslash i,p}) = \text{Auto-Differentiation}(f_{\text{KAN},i},\, x_{\backslash i,p})$
                \State $s_{i,j,p} = \text{Standardization}(\nabla f_{\text{KAN},i}(x_{\backslash i,p}))$
            \EndFor
            \State $A_{j,i} = \text{Sparsification}(\nabla f_{\text{KAN},i},\, s_{i,j,p})$
        \EndFor
        \State \textbf{return} $A$
    \end{algorithmic}
\end{algorithm}

We outline the scKAN algorithm in Algorithm~\ref{alg:scKAN}.
The algorithm takes scRNA-seq data $X\in\mathbb{R}^{g\times c}$ as input and outputs the adjacency matrix $A\in\mathbb{R}^{g\times g}$.
The algorithm trains a KAN neural network $f_i$ for each target gene $i$ with $X_{\backslash i}$ as input and $X_i$ as output.
$f_i$ serves as a predictor that reconstructs the expression of gene $i$ based on the expression of all other genes.
The gradient of the predictor $\nabla f_{\text{KAN},i}(x_{\backslash i,p})$ is calculated for each cell $p$ via automatic differentiation.
The standard score $s_{i,j,p}$ of the magnitude of the gradients across input gene dimension is computed following Eq.~(\ref{eq:standardization}). 
The adjacency matrix $A$ is updated by sparsification following Eq.~(\ref{eq:sparsification}). 
The algorithm returns the adjacency matrix $A$ as the inferred GRN.

% by Aoran
\begin{table}[!ht]
    \centering
    \caption{Comparison between scKAN and BEELINE baseline models on their properties, inferred GRN types and methodologies.
    Model categories include Correlation (Corr), ODE, Regression (Reg), GC, MI and OvR.
    }
    \label{tab:models_compare}
    \begin{small}
    \begin{tabular}{lccl} \toprule[1pt]\midrule[0.3pt]
                                             & Category           & ~Signed?~  & Description                                                                                          \\ \midrule  
    PPCOR~\cite{kim2015ppcor}                & Corr        & \centering\xmark  & \begin{minipage}[t]{0.55\textwidth}{A model that partial correlation between gene expressions as interaction strengths}\end{minipage}                                                              \\[0.6cm]
    LEAP~\cite{specht2017leap}               & Corr        & \centering\xmark & \begin{minipage}[t]{0.55\textwidth}{A model that uses the maximum Pearson correlations at different time lags as the gene interaction strength}\end{minipage}                                                    \\[1cm]
    SCODE~\cite{matsumoto2017scode}          & ODE                & \centering\cmark & \begin{minipage}[t]{0.55\textwidth}{A linear ODE model for the gene expression level}\end{minipage}                                                                                                              \\[0.6cm]
    GRISLI~\cite{aubin2020grisli}            & ODE                & \centering\xmark & \begin{minipage}[t]{0.55\textwidth}{A model similar to SCODE but with sparsity constraint and less restrictions due to the stability selection algorithm}\end{minipage}                                           \\[1cm]
    GRNVBEM~\cite{sanchez2018grnvbem}        & Reg         & \centering\cmark & \begin{minipage}[t]{0.55\textwidth}{A multivariate AR1MA1 time series model solved by the variational Bayesian inference}\end{minipage}                                                                            \\[0.6cm]
    SINCERITIES~\cite{papili2018sincerities} & Reg         & \centering\cmark & \begin{minipage}[t]{0.55\textwidth}{Models the gene distributional shift distances by linear regression to spot significant gene interactions}\end{minipage}                                                       \\[1cm]
    SINGE~\cite{deshpande2022singe}          & GC  & \centering\xmark & \begin{minipage}[t]{0.55\textwidth}{Granger causality model that ensemble vector autoregression models sparsified by generalized lasso Granger tests~\cite{bahadori2012granger}}\end{minipage}                     \\[1cm]
    PIDC~\cite{chan2017pidc}                 & MI & \centering\xmark & \begin{minipage}[t]{0.55\textwidth}{A model that decomposes partial information and quantifies the unique information between each pair of genes in all possible gene triplets.}\end{minipage} \\[1cm]
    Scribe~\cite{qiu2020scribe}              & MI & \centering\xmark & \begin{minipage}[t]{0.55\textwidth}{Causality inferencing model on GRN that computes restricted directed information~\cite{rahimzamani2016network} and its variants~\cite{rahimzamani2017potential}}\end{minipage} \\[1cm]
    GENIE3~\cite{huynh2010genie3}            & OvR               & \centering\xmark & \begin{minipage}[t]{0.55\textwidth}{An OvR model using random forests as predictors and feature importance as gene interaction strength}\end{minipage}                                                             \\[1cm]
    GRNBOOST2~\cite{moerman2019grnboost2}    & OvR               & \centering\xmark & \begin{minipage}[t]{0.55\textwidth}{An OvR model using gradient boosting machines as predictors and feature importance as gene interaction strength}\end{minipage}                                                 \\[1cm]
    scKAN (Ours)                             & OvR     & \centering\cmark & \begin{minipage}[t]{0.55\textwidth}{An OvR model using KAN neural networks as predictors and calculating gene interaction based on gradients}\end{minipage}                                                       \\ \midrule[0.3pt]\bottomrule[1pt]
    \end{tabular}
    \end{small}
\end{table}

Table~\ref{tab:models_compare} compares scKAN with BEELINE baseline models on their properties, inferred GRN types and methodologies.
We filter out the models that require prior knowledge of the GRN~\cite{woodhouse2018scns}.
We categorize the models into six categories: correlation (Corr), ordinary differential equation (ODE), Regression (Reg), Granger causality (GC), Mutual information (MI), and One-vs-Rest (OvR).
We also indicate their ability to infer GRNs with edge types (activation and inhibition).
A short description of each model is provided to give a brief overview of the model's methodology.

\newpage

\begin{figure}[!h]
    \centering
    \includegraphics[width=\textwidth]{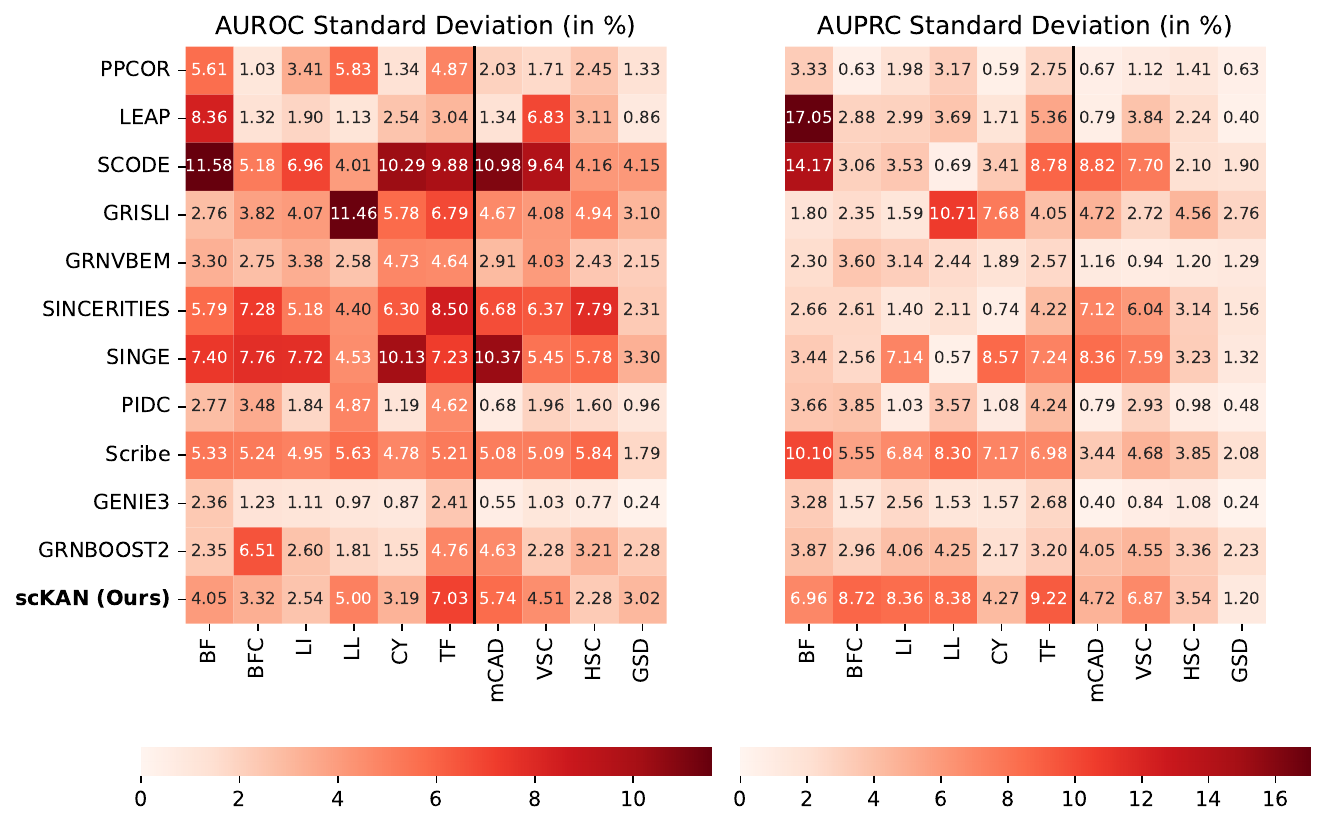}
    \caption{Standard deviation of AUROC (left) and AUPRC (right) of directed GRN inference models with unsigned ground-truth on different datasets out of 10 runs.
             Synthetic and curated datasets are separated by a vertical line.}
    \label{fig:directed_std}
\end{figure}

Fig.~\ref{fig:directed_std} shows the standard deviation of the AUROC and AUPRC values for directed GRN inference models with unsigned ground-truth.
We observed that scKAN has a stable performance across different datasets in AUROC.
scKAN has a relatively high standard deviation in AUPRC because scKAN has a much higher AUPRC value (see Section~5.2) than the average. 

\begin{figure}[!h]
    \centering
    \includegraphics[width=\textwidth]{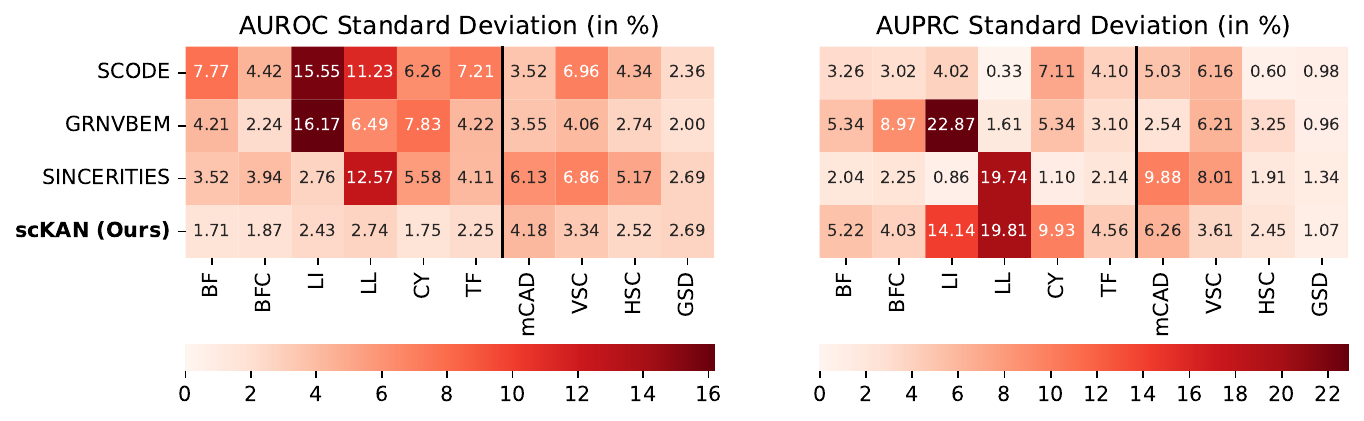}
    \caption{Standard deviation of AUROC (left) and AUPRC (right) of directed GRN inference models with signed ground-truth on different datasets out of 10 runs.
             Synthetic and curated datasets are separated by a vertical line.}
    \label{fig:signed_directed_std}
\end{figure}

Fig.~\ref{fig:signed_directed_std} shows the standard deviation of the AUROC and AUPRC values for directed GRN inference models with signed ground-truth.
We observed that scKAN has a much lower standard deviation in AUROC than other models, suggesting that current state-of-the-art models are unstable in inferring edge types, while scKAN is robust in this aspect.
Nevertheless, the standard deviation of the AUPRC values for scKAN is remarkably high, which is due to the high AUPRC values of scKAN in the LI, LL, CY, where the values are doubled compared to the rest of the models.

\begin{table}[!ht]
    \centering
    \caption{Average and standard deviation AUROC and AUPRC of directed GRN inference models with unsigned ground-truth in the scalability study out of 10 runs. The best models in each dataset and metric are boldfaced.}
    \label{tab:scalability_unsigned}
    \begin{small}
    \begin{tabular}{l|cc|cc} \toprule[1pt]\midrule[0.3pt]
        \multirow{2}{*}{Models} & \multicolumn{2}{c|}{AUROC (in \%)} & \multicolumn{2}{c}{AUPRC (in \%)} \\ \cmidrule{2-5}
        \multirow{2}{*}{} & 71 genes & 104 genes & 71 genes & 104 genes \\ \midrule
        PPCOR        & 66.14$\pm$1.42 & 60.51$\pm$1.88 &  7.96$\pm$0.31 &  4.61$\pm$0.25 \\
        LEAP         & 74.44$\pm$0.62 & 63.86$\pm$0.88 & 15.46$\pm$0.48 &  3.82$\pm$0.07 \\
        SCODE        & 51.70$\pm$2.20 & 47.98$\pm$1.84 &  3.79$\pm$0.63 &  2.13$\pm$0.31 \\
        GRNVBEM      & 54.44$\pm$1.78 & 53.89$\pm$1.06 &  3.99$\pm$0.25 &  2.54$\pm$0.15 \\
        SINCERITIES  & 49.94$\pm$3.30 & 50.30$\pm$3.96 &  3.41$\pm$3.30 &  2.25$\pm$0.35 \\
        PIDC         & 74.39$\pm$1.12 & 66.53$\pm$0.92 & 16.46$\pm$0.54 &  9.78$\pm$0.29 \\
        Scribe       & 50.78$\pm$1.59 & 50.93$\pm$1.55 &  3.60$\pm$0.33 &  2.15$\pm$0.25 \\
        GENIE3       & \textbf{77.46$\pm$0.37} & \textbf{73.53$\pm$0.24} & \textbf{18.43$\pm$0.38} & \textbf{11.78$\pm$0.35} \\
        GRNBOOST2    & 75.74$\pm$1.42 & 67.59$\pm$1.00 & 15.77$\pm$0.96 &  7.62$\pm$0.46 \\
        scKAN (Ours) & 69.65$\pm$1.33 & 69.68$\pm$1.29 & 13.92$\pm$0.63 &  7.25$\pm$0.67 \\
        \midrule[0.3pt]\bottomrule[1pt]
    \end{tabular}
    \end{small}
\end{table}

Table~\ref{tab:scalability_unsigned} shows the average and standard deviation of the AUROC and AUPRC values for directed GRN inference models with unsigned ground-truth in the scalability study.
GENIE3 shows superiority in larger datasets in both metrics.
PPCOR, LEAP, PIDC, GENIE3, GRNBOOST and scKAN perform significantly better than other models in terms of AUROC and AUPRC.
However, PPCOR, LEAP and PIDC also show a significant performance drop when scaling up to 104 genes, while all OvR models maintain stable performances across different dataset sizes.

\begin{table}[!ht]
    \centering
    \caption{Average and standard deviation AUROC and AUPRC of directed GRN inference models with signed ground-truth in the scalability study out of 10 runs. The best models in each dataset and metric are boldfaced.}
    \label{tab:scalability_signed}
    \begin{small}
    \begin{tabular}{l|cc|cc} \toprule[1pt]\midrule[0.3pt]
        \multirow{2}{*}{Models} & \multicolumn{2}{c|}{AUROC (in \%)} & \multicolumn{2}{c}{AUPRC (in \%)} \\ \cmidrule{2-5}
        \multirow{2}{*}{} & 71 genes & 104 genes & 71 genes & 104 genes \\ \midrule
        SCODE        & 45.87$\pm$1.58 & 45.42$\pm$1.86 &  1.65$\pm$0.19 &  0.99$\pm$0.08 \\
        GRNVBEM      & 48.26$\pm$2.40 & 51.14$\pm$1.57 &  1.81$\pm$0.15 &  1.23$\pm$0.07 \\
        SINCERITIES  & 55.26$\pm$1.77 & 53.20$\pm$1.95 &  2.10$\pm$0.36 &  1.23$\pm$0.11 \\
        scKAN (Ours) & \textbf{63.80$\pm$1.06} & \textbf{62.55$\pm$1.39} &  \textbf{8.09$\pm$0.42} &  \textbf{4.27$\pm$0.46} \\
        \midrule[0.3pt]\bottomrule[1pt]
    \end{tabular}
    \end{small}
\end{table}

Table~\ref{tab:scalability_signed} shows the average and standard deviation of the AUROC and AUPRC values for directed GRN inference models with signed ground-truth in the scalability study.
We observed that scKAN performs significantly better than other models in terms of AUROC and AUPRC.
Although the performance of scKAN is inferior to GENIE3 in unsigned dataset, GENIE3 is incable of inferring edge types, and scKAN is the only model that can infer edge types among all OvR models.
Besides, scKAN does not show a significant performance drop when it was scaled to a larger network or additionally inferring edge types, compared with the unsign ground-truth in Table~\ref{tab:scalability_unsigned}, demonstrating its robustness on edge type inference.

\begin{figure}[!h]
    \centering
    \includegraphics[width=\textwidth]{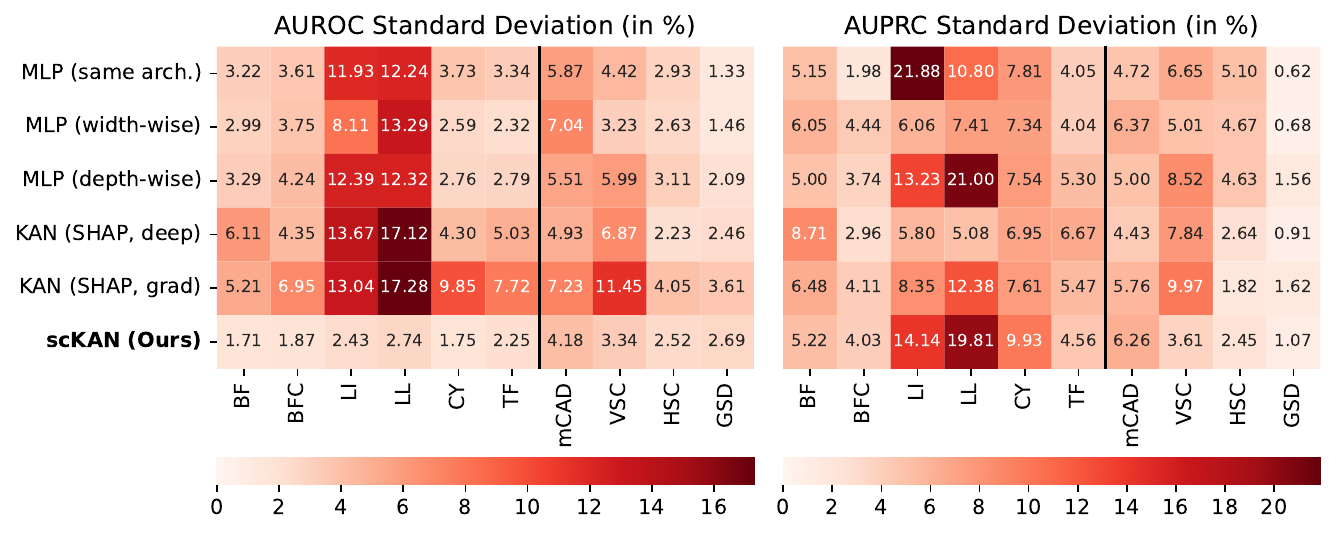}
    \caption{Standard deviation of AUROC (left) and AUPRC (right) of models in the ablation study on different datasets out of 10 runs.
             Synthetic and curated datasets are separated by a vertical line.}
    \label{fig:signed_directed_ablation_std}
\end{figure}

Fig.~\ref{fig:signed_directed_ablation_std} shows the standard deviation of the AUROC and AUPRC values for models in the ablation study.
We observed that the standard deviation of the AUROC for scKAN is the most stable, suggesting the superiority of KAN over MLP in our setting. The use of a raw gradient instead of XAI tools is sufficient to explain the model's behavior.

\begin{figure}[!h]
    \centering
    \subfigure[]{\includegraphics[width=0.48\textwidth]{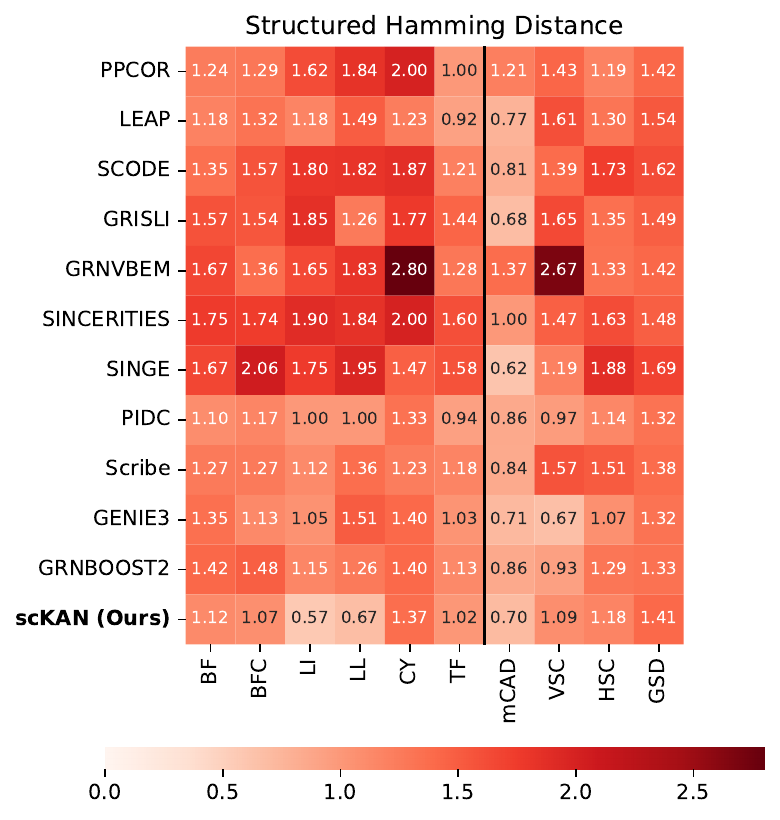}} 
    \subfigure[]{\includegraphics[width=0.48\textwidth]{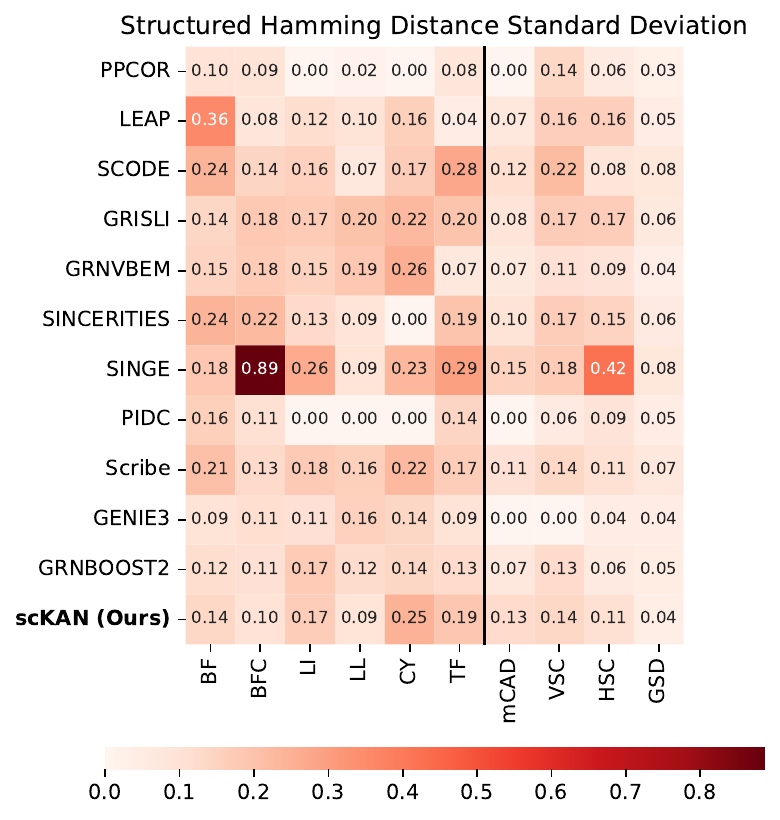}} 
    \caption{Average (a) and standard deviation (b) of the Structured Hamming Distance of directed GRN inference models with unsigned ground-truth out of 10 runs. Synthetic and curated datasets are separated by a vertical line.}
    \label{fig:directed_shd}
\end{figure}

Fig.~\ref{fig:directed_shd} shows the average and standard deviation of the Structured Hamming Distance (SHD) for directed GRN inference models with unsigned ground-truth.
All models generally performed better on the mCAD dataset, and scKAN performed exceptionally well on the LI and LL datasets.
LEAP, PIDC and OvR models have better results than the rest of the models across most of the datasets, agreeing with the conclusions made by the AUROC (Fig.~2), AUPRC (Fig.~3) and the scalability test (Table~2).

\begin{figure}[!h]
    \centering
    \subfigure[]{\includegraphics[width=0.48\textwidth]{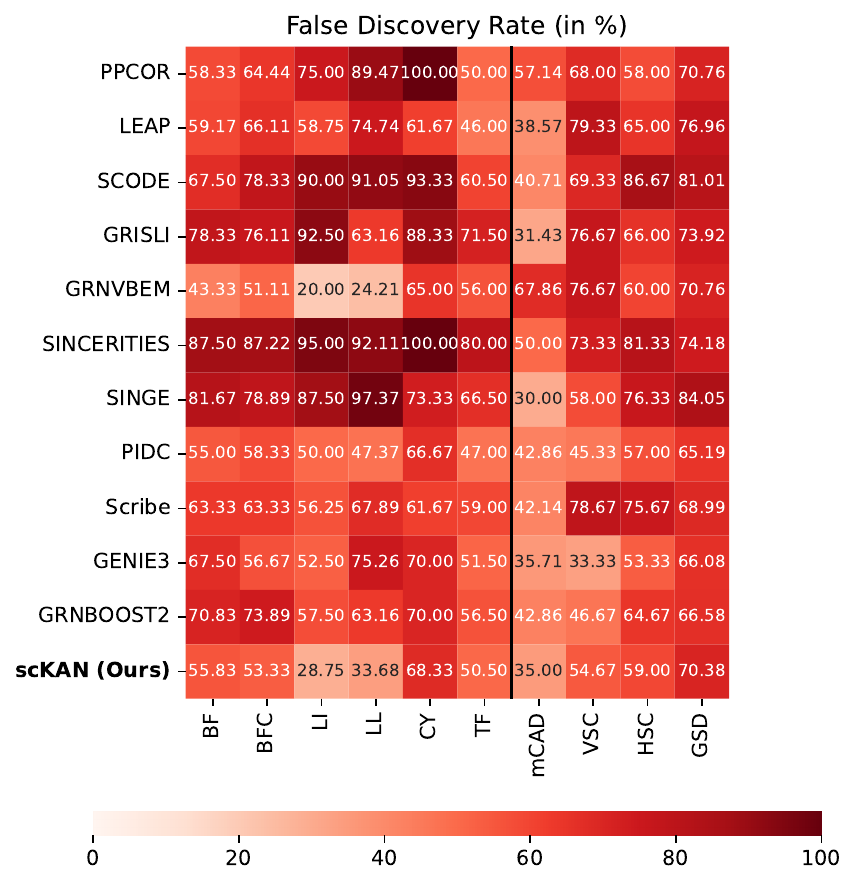}} 
    \subfigure[]{\includegraphics[width=0.48\textwidth]{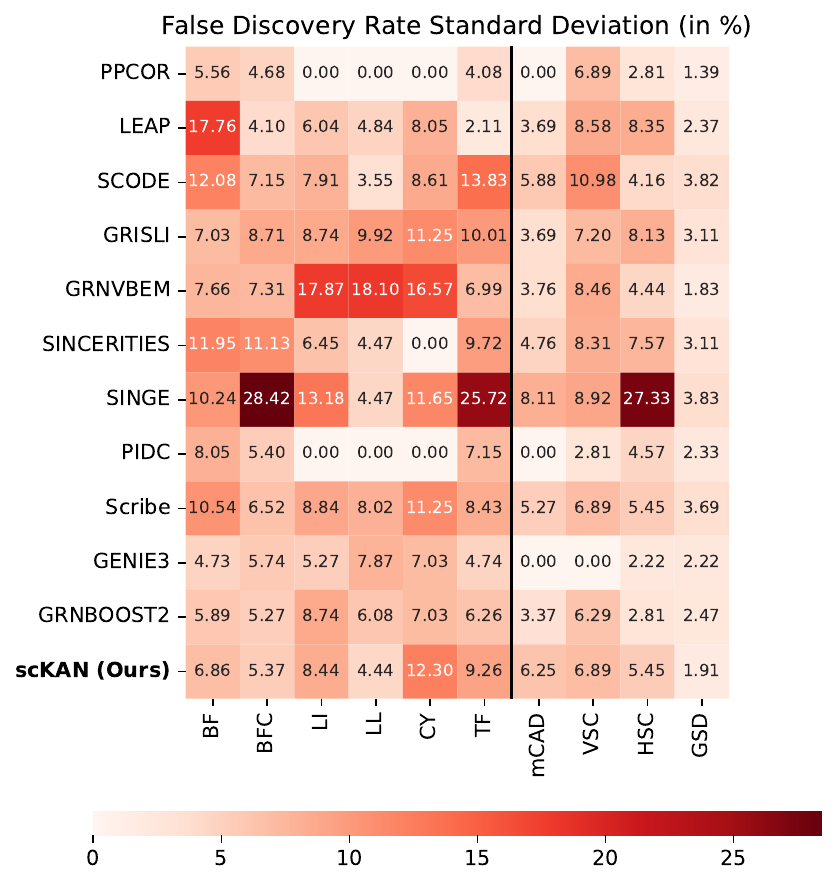}} 
    \caption{Average (a) and standard deviation (b) of the False Discovery Rate of directed GRN inference models with unsigned ground-truth out of 10 runs. Synthetic and curated datasets are separated by a vertical line.}
    \label{fig:directed_fdr}
\end{figure}

Fig.~\ref{fig:directed_fdr} shows the average and standard deviation of the False Discovery Rate (FDR) for directed GRN inference models with unsigned ground-truth.
In most of the datasets, all models have high FDR values, demonstrating the difficulty of reverse engineering GRN.
This may suggest scRNA-seq data are not sufficient to reveal the underlying GRNs.
Integrating prior knowledge and multi-omics data might be the way out to boost the GRN inference performance.

\end{document}